\documentclass{aastex631}

\graphicspath{{./}{figures/}}

\begin{document}
\title{Cloudy in the microcalorimeter era: improved energies for Si and S K$\alpha$ fluorescence lines}

\correspondingauthor{Francesco Camilloni}
\email{fcam@mpe.mpg.de}

\author[0000-0003-4161-5709]{Francesco Camilloni}
\affiliation{Dipartimento di Matematica e Fisica, Universit\`a degli Studi Roma Tre, via della Vasca Navale 84, I-00146 Roma Italy}

\author[0000-0002-4622-4240]{Stefano Bianchi}
\affiliation{Dipartimento di Matematica e Fisica, Universit\`a degli Studi Roma Tre, via della Vasca Navale 84, I-00146 Roma Italy}

\author[0000-0003-0593-4681]{Roberta Amato}
\affiliation{IRAP, Universit\'e de Toulouse, CNRS, 9 avenue du Colonel Roche, 31028 Toulouse, France}

\author[0000-0003-4503-6333]{Gary Ferland}
\affiliation{University of Kentucky, Lexington, USA}

\author[0000-0003-2538-0188]{Victoria Grinberg}
\affiliation{European Space Agency (ESA), European Space Research and Technology Centre (ESTEC), Keplerlaan 1, 2201 AZ Noordwijk, The Netherlands}

\begin{abstract}
The upcoming X-ray missions based on the microcalorimeter technology require exquisite precision in spectral simulation codes in order to match the unprecedented spectral resolution. In this work, we improve the fluorescence K$\alpha$ energies for Si \textsc{ii-xi} and S \textsc{ii-xiii} in the code Cloudy. In particular, we provide here a patch to update the Cloudy fluorescence energy table, originally based on \citet{1993A&AS...97..443K}, with the laboratory energies measured by \citet{2016ApJ...830...26H}. The new Cloudy simulations were used to model the \textit{Chandra}/HETG spectra of the High Mass X-ray Binary Vela X-1 \citep[previously presented in][]{2021A&A...648A.105A}, showing a remarkable agreement and a dramatic improvement with respect to the current release version of Cloudy (C17.02). 
\end{abstract}


\section{Introduction} \label{sec:intro}

Inner-shell ionization is responsible for some of the most important transitions in the X-ray domain. If one of the inner shell electrons of an atom or ion is hit by a photon (photoionization) or, to a lesser extent, by an electron (collisional ionization) with energy equal or higher than its ionization energy, it can be removed from the shell. The vacancy created in this way can be filled by an electron from a higher shell in two ways. The electron can lose energy either giving it to another electron (Auger transition which is radiation less) or by fluorescence, i.e. a radiative transition. The X-ray spectra of a variety of astrophysical sources are rich in fluorescence emission lines of all elements and ions of all stages. 

The launch in the early 2000s of the \textit{Chandra} and XMM-\textit{Newton} observatories provided for the first time high-resolution spectra in X-rays. The upcoming microcalorimeter-based missions are expected to represent a giant step forward, starting the era of high-precision X-ray spectroscopy. The \textit{Hitomi} mission \citep{2016SPIE.9905E..0UT} demonstrated the breakthrough capabilities of this technology \citep[see e.g.][]{2019MNRAS.483.1701S}. The next X-ray missions to be launched with microcalorimeters on board, \textit{XRISM} \citep{2018SPIE10699E..22T} and \textit{Athena} \citep{2013sf2a.conf..447B}, will have an energy resolution of a few eV in all the X-ray band, coupled with a large effective area. The analysis and interpretation of the X-ray spectra provided by these new missions will present unprecedented challenges. 

Therefore, we have started a process to update the spectral simulation code Cloudy \citep{2017RMxAA..53..385F}, in order to keep up with the spectroscopic requirements of these new X-ray missions \citep{Chakraborty_2020a,Chakraborty_2020b,Chakraborty_2021}. In this work, we present a first attempt to update the  \cite{1993A&AS...97..443K} database, used by Cloudy for fluorescence emission. In particular, we consider the experimental data for fluorescent K$\alpha$ energies for Si \textsc{ii-xi} and S \textsc{ii-xiii} taken by \cite{2016ApJ...830...26H}.

\section{Results} \label{sec:update}

\subsection{Cloudy update}

The current version of Cloudy, C17.02, uses Table 3 of \cite{1993A&AS...97..443K} as the main database for fluorescence yields, energies and number of ejected Auger electrons for all elements and ions from Be to Zn. Their calculation were in reasonable agreement with more detailed computations available at that time, but they are now unsuitable for current and future high-resolution spectroscopy.

For the update reported in this work, we decided to use the experimental data reported in \citet{2016ApJ...830...26H}. In particular, K$\alpha$ line energies\footnote{K$\alpha_{1}$ and K$\alpha_{2}$ line energies are not resolved in the experimental data reported by \citet{2016ApJ...830...26H}, so they are assumed to have the same energy. For Si and S, their difference is still below the resolution of \textit{Athena}.} from O-like to Be-like Si and S (i.e. Si \textsc{vii-xi} and S \textsc{ix-xiii}) were taken from their Table 3, which gives centroids for unresolved blends. On the other hand, for lower ionization values, the line energies are taken from Table 5, where individual values are listed for
Si \textsc{ii-iv} and Si \textsc{v-vi} and S \textsc{ii-vi} and S \textsc{vii-viii}.

\subsection{An applied case: Vela X-1}

\begin{figure}
    \centering
    \includegraphics[scale=0.8]{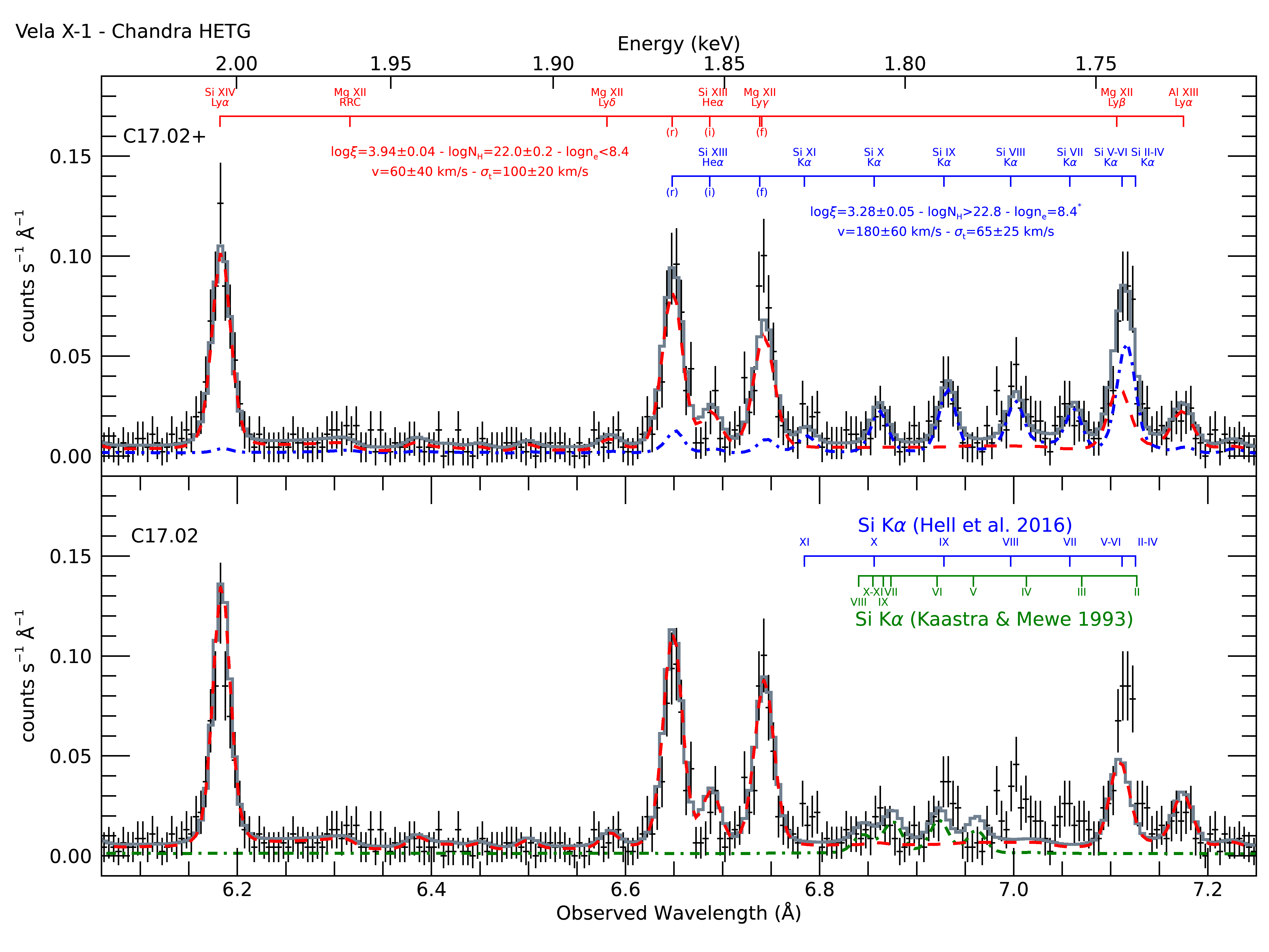}
    \caption{Visually co-added MEG $\pm1$ order spectra of the HMXB Vela X-1 at the orbital phase $\phi_{orb}=0.75$ \citep[see][]{2021A&A...648A.105A}. \textit{Top}: Best fit model with Cloudy (grey solid line), using the improved energies for the Si fluorescence lines, as described in this work (dubbed C17.02+). The specific contributions of each gas component are labelled (red dashed line and blue dot-dashed line), together with the best-fit parameters and 90\% confidence level uncertainties (see text for details). \textit{Bottom}: As above, but with the current version of Cloudy, C17.02. The low ionization component is here labelled in green, together with the adopted Si K$\alpha$ lines \citep[from][]{1993A&AS...97..443K}. For ease of comparison, the improved energies from \citet{2016ApJ...830...26H} are in blue, as in the top panel.}
    \label{fig:Vela_update_2panel}
\end{figure}

We used our updated version of Cloudy to compute an Xspec \citep{1996ASPC..101...17A} tabulated additive model  to model the emission line spectrum of the High Mass X-ray Binary Vela X-1. In particular, we applied the model to the \textit{Chandra}/HETG spectrum of Vela X-1 at the orbital phase $\phi_{orb}=0.75$ \citep[ObsID 14654,  see][for details on the data reduction and the input spectrum for the Cloudy simulations]{2021A&A...648A.105A}. Here, we restrict the fit to the 6.1-7.2 \AA\ band, where Si lines are present. The resulting best fit is shown in Fig.~\ref{fig:Vela_update_2panel} (top panel). The spectrum is very well modelled by two gas components, with different parameters\footnote{The complete Xspec model is \texttt{vashift*gsmooth*atable\string{Cloudy.fits\string} + gsmooth*vashift*atable\string{Cloudy.fits\string}}, where the \texttt{gsmooth} velocity is self-consistently linked to the turbulence velocity of the Cloudy additive model.}: the higher-ionization component (in red) produces the Si \textsc{xiii} and Si \textsc{xiv} recombination lines, while the lower-ionization component (in blue) mostly reproduces the Si fluorescence lines. The latter, whose energies were updated in this work, nicely match the observed data.

For comparison, the bottom panel of Fig.~\ref{fig:Vela_update_2panel} shows the same data, with the same model, but with Xspec tables produced with Cloudy C17.02. While the larger ionization gas component is roughly unchanged, the lower ionization component is basically unconstrained, and most of the Si fluorescence lines are not reproduced by the Cloudy additive model. This is due to the inaccurate line energies taken from  \citet{1993A&AS...97..443K} (in green), significantly different from the values reported in \citet{2016ApJ...830...26H} (in blue).

Though attempts to model the accreting wind of Vela X-1 with a multi-component plasma have been made in the past \citep[see, e.g.,][]{Lomaeva2020,2021A&A...648A.105A}, this is the first time where two contributions are clearly distinguished. At the specific orbital phase accounted in this work, the neutron star (NS) in the binary system is about to enter the eclipsing phase, moving further along the line of sight. The observer has, hence, a privileged view on the medium which has just been perturbed by the passage of the NS and photoionised by the X-ray radiation coming from its surface, the so called photoionization wake. The different ionization parameters ($\log\xi_{1}$/erg cm s$^{-1}=3.94 \pm 0.04$ and $\log\xi_{2}$/erg cm s$^{-1}=3.28 \pm 0.05$), as well as  turbulence ($\sigma_{t1}=100 \pm 20$ and $\sigma_{t2}=65 \pm 25$ km s$^{-1}$) and bulk velocities ($v_1=60\pm40$ and $v_2=180\pm60$ km s$^{-1}$), clearly point to the coexistence of two media, with different ionization and kinematic properties, very likely given by the photoionization wake embedded in the surrounding wind.

\section{Conclusions}\label{sec:conclusions}

We presented an update of the Cloudy fluorescence energy table, originally based on \citet{1993A&AS...97..443K}, with the Si and S laboratory energies measured by \citet{2016ApJ...830...26H}. The update can be applied to the current release version of Cloudy (C17.02) via the patch `Camilloni2021KMupdate.diff', which is posted to the Cloudy user group\footnote{\url{https://cloudyastrophysics.groups.io/g/Main/topics}}.

This work should be considered as a pathfinder to demonstrate the urgent need for a systematic update of the fluorescence line energies in Cloudy for all elements and ions, since the inaccurate values in C17.02 already affect the modelling of current gratings spectra, and will soon become obsolete with the advent of microcalorimeter based X-ray missions.

\begin{acknowledgments}
SB acknowledges support by EU H2020 (grant  871158), ASI (grant 2017-12-H.0) and PRIN MIUR (project 2017-PH3WAT).
GJF acknowledges support by NSF (1816537, 1910687), NASA (ATP 17-ATP17-0141, 19-ATP19-0188), and STScI (HST-AR- 15018 and HST-GO-16196.003-A).

\end{acknowledgments}
\newpage
\bibliography{RNVelaX1}
\bibliographystyle{aasjournal}

\end{document}